\documentclass{article}

\PassOptionsToPackage{numbers, compress}{natbib}
\usepackage[final]{neurips_2021_ml4ps}

\usepackage[utf8]{inputenc} 
\usepackage[T1]{fontenc}    
\usepackage{url}            
\usepackage{booktabs}       
\usepackage{graphicx}
\usepackage{amsfonts}       
\usepackage{nicefrac}       
\usepackage{microtype}      
\usepackage{xcolor}         
\usepackage{amsmath}
\usepackage{wrapfig}
\usepackage{hyperref}       

\title{A deep ensemble approach to X-ray polarimetry}

\author{%
  A.L. Peirson\thanks{\url{https://www.alpeirson.com}} , R.W. Romani \\
  Kavli Institute for Particle Astrophysics and Cosmology\\
  Stanford University\\
  \texttt{alpv95@stanford.edu} \\
}

\begin{document}

\maketitle

\begin{abstract}
X-ray polarimetry will soon open a new window on the high energy universe with the launch of NASA's Imaging X-ray Polarimetry Explorer (IXPE). Polarimeters are currently limited by their track reconstruction algorithms, which typically use linear estimators and do not consider individual event quality. 
We present a modern deep learning method for maximizing the sensitivity of X-ray telescopic observations with imaging polarimeters, with a focus on the gas pixel detectors (GPDs) to be flown on IXPE. 
We use a weighted maximum likelihood combination of predictions from a deep ensemble of ResNets, trained on Monte Carlo event simulations. We derive and apply the optimal event weighting for maximizing the polarization signal-to-noise ratio (SNR) in track reconstruction algorithms. For typical power-law source spectra, our method improves on the current state of the art, providing a $\sim40$\% decrease in required exposure times for a given SNR. 
\end{abstract}

\section{Introduction}
Measuring X-ray polarization, the degree of order in X-ray electric field oscillations, has been a major goal in astrophysics for the last 50 years. X-ray polarization measurements offer rich opportunities to probe the magnetic field topology and emission physics of high energy astrophysical sources, such as accreting black holes and astrophysical jets \cite{krawczynski_using_2019, weisskopf_overview_2018}. 
The recent development of photoelectron tracking detectors \cite{bellazzini_novel_2003} has greatly improved the prospects of doing so. The gas pixel detector (GPD) \cite{bellazzini_sealed_2007} has brought soft X-ray polarimetry (1-10 keV) to the PolarLight CubeSat test \cite{feng_x-ray_2020}, the scheduled NASA IXPE mission \cite{sgro_imaging_2019}, and the potential Chinese mission, eXTP \cite{zhang_extp_2017}.

Imaging X-ray polarization telescopes feed GPDs which directly image charge tracks formed from photoelectrons scattered by incoming X-ray photons. 
IXPE \cite[planned for launch December 9th 2021]{weisskopf_overview_2018,odell_imaging_2018} will use three co-aligned X-ray telescopes, whose focal planes are imaged by GPDs with hexagonal pixels.
IXPE's sensitivity is limited by the track analysis algorithm used to recover source polarization, spatial structure and energy, given a measured set of electron track images. 
In the $1-10$ keV range, the cross-section for photoelectron emission is proportional to cos$^2(\theta)$, where $\theta$ is the angle between the normal incidence X-ray's electric vector position angle (EVPA) and the azimuthal emission direction of the photoelectron. By measuring a large number of individual photoelectron emission angles $\theta$, one can recover the above distribution to extract the source polarization parameters: polarization fraction ($0 \leq p_0 \leq 1$) and electric vector position angle (EVPA, $-\pi/2 \leq \phi < \pi/2$). In practice, the recovery of photoelectron emission angles from track images is imperfect. Track images are noisy due to Coulomb scattering and diffusion, and, especially for low energies, are often barely resolved. Emission angle estimates are highly heteroskedastic.

The current track reconstruction method for GPDs is a moment analysis described by \cite{bellazzini_novel_2003}. Impressive accuracies are achieved from a simple weighted combination of track moments. However, simple moments cannot capture all of the image information, especially for long high energy tracks, and so a more sophisticated image analysis scheme may lead to improved track angle recovery. 
The problem of recovering polarization parameters from a dataset of (IXPE) electron track images has recently been announced as an open problem in the machine learning community \cite{moriakov_inferring_2020}.

In this work, we aim to both improve individual event estimates and explicitly model their uncertainty to increase final polarization sensitivity.
We demonstrate a two step method: (1) Use a deep ensemble of ResNets \cite{he_deep_2015} to predict electron track angles $\theta$ and their uncertainties. (2) Combine predicted angles and their uncertainties in a weighted estimator of the polarization parameters $p_0, \phi$ that maximizes the SNR. 
Our empirical findings indicate a substantial improvement over the current state of-the-art track reconstruction \cite{bellazzini_novel_2003, kitaguchi_convolutional_2019-1}. 
While the results shown here are specific to IXPE's GPDs, the methods are general, and can be applied to other imaging detector geometries. Extended results, including spatial and energy resolution are discussed in \cite{peirson_deep_2021,peirson_towards_2021}.

\section{Step I: deep ensemble}
\label{sec:deep}
\begin{figure*}
\centering
\includegraphics[width=0.8\textwidth]{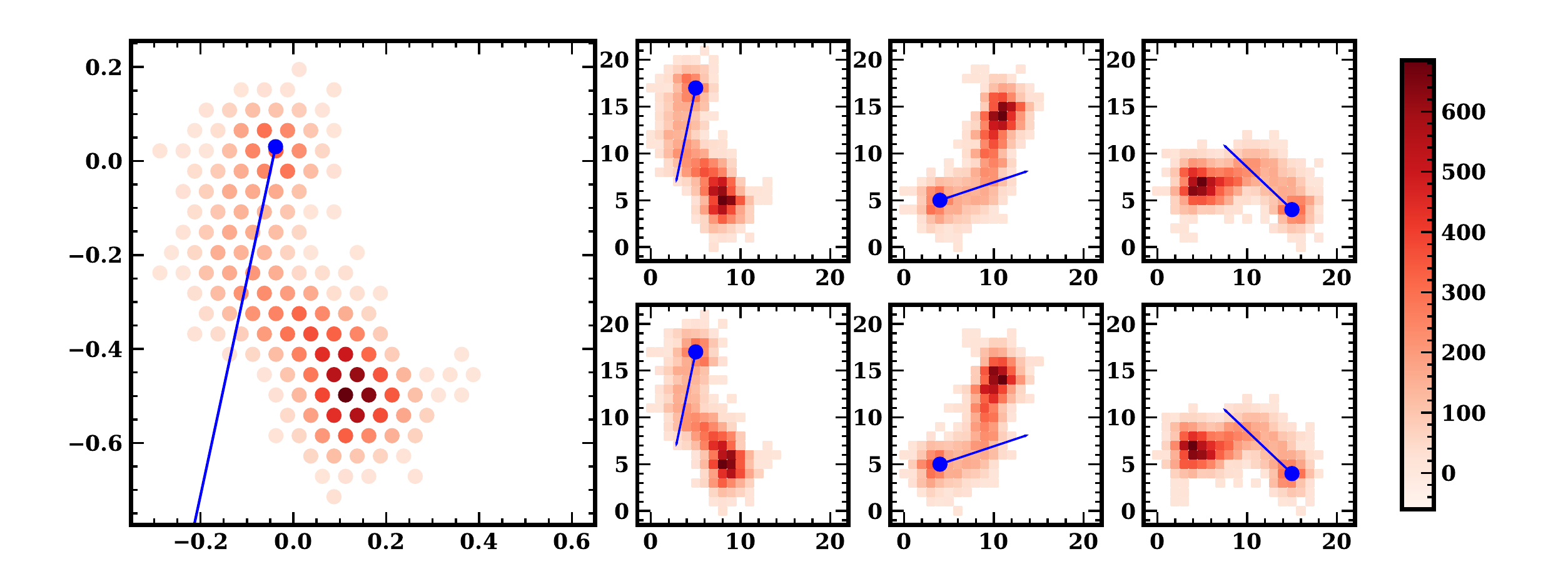}
\caption{Example square conversions of a 6.4 keV hexagonal track (left panel). The six panels to the right show shifts along the 120$^\circ$ GPD axes; shifting odd rows (upper) or even rows (lower). For each hexagonal track, the ResNet-18s are fed column-wise pairs of square conversions, along with the initial photoelectron direction (blue line) as a label.}
\label{fig:square}
\end{figure*}
To extract the angles from individual tracks we use a supervised deep learning technique known as deep ensembles \cite{lakshminarayanan_simple_2017}. Deep ensembles are made up of an ensemble of individual neural network models, each trained independently on the same data set to predict the desired output features.
Deep ensembles provide estimates of the predictive uncertainty by exploiting different random initializations of the same model at the start of training which leads to widely different prediction functions \cite{fort_deep_2019}. In terms of performance and scalability, deep ensembles remain the current state of the art in uncertainty quantification \cite{hoffmann_deep_2021}.
In our case, we have an image to feature regression problem. Convolutional neural networks have been designed with a spatial inductive bias appropriate for image regression problems. So our deep ensemble will be made up of $M$ individual ResNet-18s \cite{he_deep_2015}.

To make the hexagonal track images admissable inputs to standard ResNet architectures, we first convert the hexagonal images to square image arrays by shifting every other column and rescaling the distance between points, as described in \cite{steppa_hexagdly_2019}. Since there are two possible shifts (odd and even rows), we apply both and stack the two shifted images, similar to color channels in $rgb$ images. We do this to more closely approximate spatial equivariance of the ResNet convolution kernels in the hexagonal space. At test time, we apply the deep ensemble to the same track 3 times, each time rotated by $120^{\circ}$ in hexagonal space. We find this reduces all relevant prediction bias on $\theta$ (and later $p_0, \phi$) introduced when converting from hexagonal to square coordinates. Fig.~\ref{fig:square} summarizes this process. An alternative solution is to use a model with native hexagonal convolutions, such as \cite{steppa_hexagdly_2019}. In practice, we found it more effective and expedient to leverage existing square convolution architectures.

For part (1) of our method, we need to predict emission angle $\theta$ and its associated uncertainty for each track. Instead of the Gaussian negative log-likelihood (NLL) used in \cite{lakshminarayanan_simple_2017}, we use the von Mises (VM) distribution NLL as our loss function, the maximum entropy distribution for circular data with specified expectation value. This more appropriately reflects the distribution of our estimators $\hat{\theta}$, which are clearly periodic.
The VM distribution is parameterized by concentration parameter $\kappa$; for large $\kappa$ the VM converges to a Gaussian with variance $\sigma^2 = 1/\kappa$ and for small $\kappa$ it approaches a uniform distribution.
We parameterize the true angle $\theta$ as a 2D vector $\mathbf{v} = (\rm{cos}2\theta,\rm{sin}2\theta)$ to capture the periodicity. Only $-\pi/2 \leq \hat{\theta} < \pi/2$, as opposed to $-\pi \leq \hat{\theta} < \pi$, is required for polarization estimation since $-\pi/2 \leq \phi < \pi/2$; see eq.~\ref{eq:likelihood}). 
Each of the ResNet-18 models in our ensemble computes the loss as
\begin{equation}
    \rm{Loss}(\hat{\mathbf{v}}, \hat{\kappa} | \mathbf{v}) \propto -\hat{\kappa}(\hat{\mathbf{v}}.\mathbf{v}) + \log I_0(\hat{\kappa}) .
    \label{eqn:loss}
\end{equation}
where $I_0$ is the modified Bessel function of the first kind.
Each ResNet-18 in the ensemble ($j = 1:M$) outputs the 3-vector $(\mathbf{\hat{v}}_{ij}, \hat{\kappa}_{ij})$ for track $i$. Then $\hat{\theta}_{ij} = \rm{arctan}\big(\frac{\hat{v}_{{ij}_2}}{\hat{v}_{{ij}_1}}\big) / 2$ and $\hat{\kappa}_{ij}$ defines the predicted aleatoric uncertainty. The final track angle prediction $\hat{\theta}_i$ is the circular average over the ensemble predictions $j = 1:M$. The epistemic uncertainty on $\hat{\theta}_i$ is also assumed to follow a VM distribution, with concentration parameter $\kappa^e_i$; we estimate it using the appropriate maximum likelihood estimator for $\hat{\kappa}^e_i$ given the independent sample set $\{\theta_{ij}\}^M_{j=1}$, see appendix.
The total predictive error on each track angle $\hat{\theta}_i$ is given by summing the aleatoric and epistemic variances: $1/\hat{\kappa}_i^{tot} = (1/M\sum_j1/\hat{\kappa}_{ij})  + 1/\hat{\kappa}^e_{i}$

\paragraph{Data.} Our intial dataset consists of 3.5 million GEANT4 \cite{agostinelli_geant4simulation_2003} Monte Carlo simulated tracks, where each track is labelled with its 2D emission angle vector, an example of which is shown in fig.~\ref{fig:square}. The track energies uniformly span $1.0 - 10.0$ keV, IXPE's most sensitive range, and are unpolarized (uniform track angle distribution).
Since we don't know the true event energy, we want a model that can make predictions for tracks of all energies.
We have confirmed the simulated track data matches real flight detector data to extremely high precision, and that our method is robust to any remaining covariate shift, maintaining the relative improvement shown in \S4 for real detector data. 

\paragraph{Training.} 
We apply pixelwise normalization to the square track images. 
Each ResNet-18 model is trained with a Momentum Optimizer, and a step-wise learning rate beginning at $1 \times 10^{-3}$ and ending at $1 \times 10^{-5}$ on 2 NVIDIA GeForce RTX 2080 GPUs.
ResNets with batch size $512,1024,2048$ are all considered for selection in the final ensemble. We randomly select $M=10$ trained ResNets to compose the final ensemble, whose distributed results estimate the epistemic error.

\section{Step II: polarization estimation}
The basic problem is to estimate $p_0$ and $\phi$ from a set of measured track angles $\{\hat{\theta}_i\}_{i=1}^N$. As described in the introduction, true track angles exhibit a sinusoidal modulation with period $\pi$
\begin{equation} \label{eq:likelihood}
    p(\theta|p_0,\phi) = \frac{1}{2\pi}\big(1 + p_0{\rm cos}\big[2(\theta - \phi)\big]\big)
\end{equation}
where $0 \leq p_0 \leq 1$, $-\pi/2 \leq \phi < \pi/2$ and $-\pi \leq \theta < \pi$. One could estimate $(p_0, \phi)$ using the maximum likelihood estimator (MLE) for eq.\ref{eq:likelihood}. Equivalently, \cite{kislat_analyzing_2015} have shown that the minimum variance unbiased estimator for $(p_0, \phi)$ is given by
\begin{equation}
    \hat{p}_0 = \sqrt{\hat{\mathcal{Q}}^2 + \hat{\mathcal{U}}^2},
    \label{eqn:p}
\end{equation}
\begin{equation}
    \hat{\phi} = \frac{1}{2}\arctan\frac{\hat{\mathcal{U}}}{\hat{\mathcal{Q}}}.
    \label{eqn:th}
\end{equation}
where $\hat{I} = N$, $\hat{\mathcal{Q}} = \frac{1}{\hat{I}}\sum_{i=1}^N2\cos2\hat{\theta}_i,$ and  $\hat{\mathcal{U}} = \frac{1}{\hat{I}}\sum_{i=1}^N2\sin2\hat{\theta}_i$. However, since $\hat{\theta}_i$ are imperfectly recovered, we adjust eq.~\ref{eq:likelihood} to include the VM uncertainty on our estimates, $\hat{\theta} = \theta + {\rm VM}(0,\kappa)$. Computing the convolution of eq.~\ref{eq:likelihood} with the VM, we find:
\begin{equation} \label{eq:likelihood2}
    p(\hat{\theta}|p_0,\phi) = \frac{1}{2\pi}\big(1 + p_0\frac{I_1(\kappa)}{I_0(\kappa)}{\rm cos}\big[2(\hat{\theta} - \phi)\big]\big)
\end{equation}
In other words, $\hat{\theta}$ follow the same distribution as $\theta$ but smeared by a modulation factor $\mu = {I_1(\kappa)}/{I_0(\kappa)}$. Current analyses \cite{kitaguchi_convolutional_2019-1, moriakov_inferring_2020} treat $\mu$ as constant for all tracks and calculate it based on broadband calibration measurements: no connection to individual track predictions. In effect, they assume homoskedastic emission angle measurements.
Here, we explicitly model the heteroskedasticity of our predictions $\hat{\theta}$ by using a deep ensemble  (\S\ref{sec:deep}) to estimate event uncertainties $\hat{\kappa}$ and include them in the likelihood for final $(p_0,\phi)$ predictions. 

Instead of explicitly maximizing a likelihood function based on eq.\ref{eq:likelihood2} and our predictions $\{\hat{\theta}_i,\hat{\kappa}_i\}^{N}_{i=1}$ to estimate $(p_0, \phi)$, we can define event weights 
\begin{equation}
w_i = {I_1(\hat{\kappa}_i)}/{I_0(\hat{\kappa}_i)},
\end{equation}
and use eq.\ref{eqn:p}-\ref{eqn:th} by defining 
\begin{equation}
\label{eqn:neffeq}
    \hat{I} = N_{\rm eff} = \frac{(\sum_{i=1}^N w_i)^2}{\sum_{i=1}^N w_i^2}, 
\end{equation}
\begin{equation}
    \hat{\mathcal{Q}} = \frac{1}{\hat{I}}\sum_{i=1}^N2w_i\cos2\hat{\theta}_i,
\end{equation}
\begin{equation}
    \hat{\mathcal{U}} = \frac{1}{\hat{I}}\sum_{i=1}^N2w_i\sin2\hat{\theta}_i.
\end{equation}
As N approaches infinity, this converges to the MLE \cite{kislat_analyzing_2015}. 
The posterior distribution $p(p_0, \phi | \hat{p}_0, \hat{\phi})$ is given in appendix \S\ref{sec:posterior}.  




\begin{figure}[t]
\centering
\includegraphics[width=0.45\textwidth]{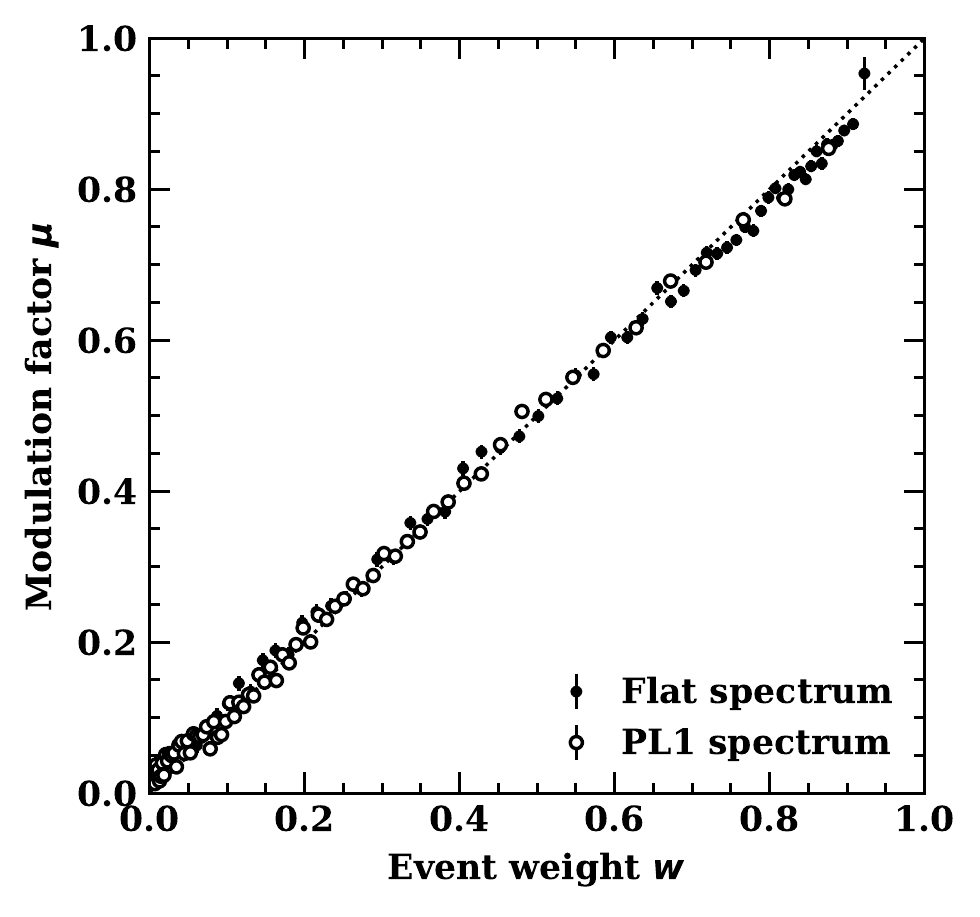}
\caption{Measured modulation factor $\mu$ as a function of predicted track weight $w = {I_1(\hat{\kappa})}/{I_0(\hat{\kappa})}$ for large test data sets of $1-10$ keV simulated events. Each $\mu$ bin is calculated from 20,000 individual track events. Open circles represent a PL1 source, closed circles a flat spectrum.}
\label{fig:errs}
\end{figure}

\begin{table}
\centering
\caption{Sensitivity analysis for $10^5$ 2-8\,keV photons with a $dN/dE \sim E^{-1}$ spectrum.}
\begin{tabular}{@{} l l @{}}
\toprule
{Method}&MDP$_{99}$(\%) \\
\midrule
{Moment Analysis} & {4.45 $\pm$ 0.02}\\ 
{Deep ensemble} &{4.21 $\pm$ 0.02} \\ 
{Deep ensemble w/ weights} &{\textbf{3.38 $\pm$ 0.01}}\\
 \bottomrule
\end{tabular}
\label{tab:fom2}
\end{table}

\paragraph{Figure of merit.} We require a figure-of-merit to define the quality of polarization reconstruction using different methods.
Aligned with prior work, we adopt the standard figure-of-merit used in X-ray polarimetry: minimum detectable polarization (MDP) \cite{weisskopf_understanding_2010}. MDP$_{99}$ is the polarization fraction that has a 1\% probability of being exceeded by chance for an unpolarized ($p_0 = 0$) source. This can be found by integrating the posterior distribution (see appendix): ${\rm MDP}_{99} = 4.29/{(\mu\sqrt{N})}$ for an unweighted polarization estimate. Here $\mu$ is the empirical modulation factor for the entire set of track observations (i.e. $\mu = \hat{p}_0$ for $p_0 = 1$). For a weighted estimate $N$ is replaced with $N_{\rm eff}$ (eq.~\ref{eqn:neffeq}) \cite{kislat_analyzing_2015}. MDP$_{99}$ is effectively a (inverse) ratio of recovered signal $\mu$ to noise $\sim 1 / \sqrt{N}$. For a given detector, track reconstruction algorithms with lower MDP$_{99}$ are better. In the appendix, we prove our chosen weighting minimizes the MDP$_{99}$ (maximizes SNR).

\section{Results and discussion}

We evaluate our two step method and the standard moment analysis on a power law 1 dataset with $N = 10^5$ tracks. Once emission angles are estimated, all methods use eqs.\ref{eqn:p}-\ref{eqn:th} for polarization prediction.
To further strengthen our analysis, we ablate using predicted uncertainties weights.
Table~\ref{tab:fom2} shows the results. 
Our weighted approach using deep ensemble predictions yields significant improvement over the standard analysis, with a  $1 - 3.38/4.45 \sim 24\%$ decrease in MDP$_{99}$. The equivalent decrease in IXPE's required exposure times to reach the same SNR is $1 - (3.38/4.45)^2 \sim 40\%$.
Our method without any uncertainty weighting provides a significantly smaller improvement. 
The gain in sensitivity of our method comes in a small part from an improvement in $\hat{\theta}$ accuracy, and in a larger part from properly modelling $\hat{\theta}$ heteroskedasticity, using deep ensemble predicted uncertainties as appropriate event weights. 
In fig.~\ref{fig:errs} we assess the trustworthiness of our deep ensemble predicted VM uncertainties. We find the empirical modulation factor matches the deep ensemble predicted modulation factor (event weights) very closely for a selection of test spectra. 
In the appendix, we visualize our deep ensemble $\hat{\theta}$ at multiple energies to assess any prediction bias.

Our results on real (non-simulated) detector data indicates similar exposure time reduction. A decrease in exposure time of $40\%$ adds public value since NASA-funded projects like IXPE can observe significantly more sources, at better SNR, over the mission lifetime.
We expect our approach of using deep ensemble predicted uncertainties to take into account heteroskedasticity could be applied to other problems in engineering and the physical sciences, for example in high energy particle physics: fitting Cauchy distributions to the frequencies of noisy (deep learning) measured decay states.

\acksection
We would like to thank Christopher Fifty for a thorough read-through of the manuscript and the referees for helpful and constructive comments. This work was supported in part by the NASA FINESST program (grant 80NSSC19K1407) and grant
NNM17AA26C from the Marshall Space Flight Center.

\bibliographystyle{natbibsty}
\bibliography{references.bib}

\begin{thebibliography}{10}

\bibitem{krawczynski_using_2019}
Krawczynski, H., Matt, G., Ingram, A.R., Taverna, R., Turolla, R. et~al.
\newblock Using {X}-{Ray} {Polarimetry} to {Probe} the {Physics} of {Black}
  {Holes} and {Neutron} {Stars}.
\newblock 51:150, May 2019.
\newblock Conference Name: Bulletin of the American Astronomical Society.

\bibitem{weisskopf_overview_2018}
Weisskopf, M.
\newblock An {Overview} of {X}-{Ray} {Polarimetry} of {Astronomical} {Sources}.
\newblock {\em Galaxies}, 6:33, March 2018.

\bibitem{bellazzini_novel_2003}
Bellazzini, R., Angelini, F., Baldini, L., Brez, A., Costa, E. et~al.
\newblock Novel gaseous x-ray polarimeter: data analysis and simulation.
\newblock In {\em Polarimetry in {Astronomy}}, volume 4843, pages 383--393.
  International Society for Optics and Photonics, February 2003.

\bibitem{bellazzini_sealed_2007}
Bellazzini, R., Spandre, G., Minuti, M., Baldini, L., Brez, A. et~al.
\newblock A sealed {Gas} {Pixel} {Detector} for {X}-ray astronomy.
\newblock {\em Nuclear Instruments and Methods in Physics Research A},
  579:853--858, September 2007.

\bibitem{feng_x-ray_2020}
Feng, H. and Bellazzini, R.
\newblock The {X}-ray polarimetry window reopens.
\newblock {\em Nature Astronomy}, 4(5):547--547, May 2020.
\newblock Number: 5 Publisher: Nature Publishing Group.

\bibitem{sgro_imaging_2019}
Sgrò, C. and {IXPE Team}.
\newblock The {Imaging} {X}-ray {Polarimetry} {Explorer} ({IXPE}).
\newblock {\em Nuclear Instruments and Methods in Physics Research A},
  936:212--215, August 2019.

\bibitem{zhang_extp_2017}
Zhang, S.N., Feroci, M., Santangelo, A., Dong, Y.W., Feng, H. et~al.
\newblock {eXTP}: {Enhanced} {X}-ray {Timing} and {Polarization} mission.
\newblock In {\em Space {Telescopes} and {Instrumentation} 2016: {Ultraviolet}
  to {Gamma} {Ray}}, volume 9905, page 99051Q. International Society for Optics
  and Photonics, January 2017.

\bibitem{odell_imaging_2018}
O'Dell, S.L., Baldini, L., Bellazzini, R., Costa, E., Elsner, R.F. et~al.
\newblock The {Imaging} {X}-ray {Polarimetry} {Explorer} ({IXPE}): technical
  overview.
\newblock 0699:106991X, August 2018.
\newblock Conference Name: Space Telescopes and Instrumentation 2018:
  Ultraviolet to Gamma Ray.

\bibitem{moriakov_inferring_2020}
Moriakov, N., Samudre, A., Negro, M., Gieseke, F., Otten, S. and Hendriks, L.
\newblock Inferring astrophysical {X}-ray polarization with deep learning.
\newblock {\em arXiv:2005.08126 [astro-ph]}, May 2020.
\newblock arXiv: 2005.08126.

\bibitem{he_deep_2015}
He, K., Zhang, X., Ren, S. and Sun, J.
\newblock Deep {Residual} {Learning} for {Image} {Recognition}.
\newblock {\em arXiv:1512.03385 [cs]}, December 2015.
\newblock arXiv: 1512.03385.

\bibitem{kitaguchi_convolutional_2019-1}
Kitaguchi, T., Black, K., Enoto, T., Hayato, A., Hill, J.E., Iwakiri, W.B.,
  Kaaret, P., Mizuno, T. and Tamagawa, T.
\newblock A convolutional neural network approach for reconstructing
  polarization information of photoelectric {X}-ray polarimeters.
\newblock {\em Nuclear Instruments and Methods in Physics Research Section A:
  Accelerators, Spectrometers, Detectors and Associated Equipment}, 942:162389,
  October 2019.
\newblock arXiv: 1907.06442.

\bibitem{peirson_deep_2021}
Peirson, A.L., Romani, R.W., Marshall, H.L., Steiner, J.F. and Baldini, L.
\newblock Deep ensemble analysis for {Imaging} {X}-ray {Polarimetry}.
\newblock {\em Nuclear Instruments and Methods in Physics Research Section A:
  Accelerators, Spectrometers, Detectors and Associated Equipment}, 986:164740,
  January 2021.

\bibitem{peirson_towards_2021}
Peirson, A.L. and Romani, R.W.
\newblock Towards {Optimal} {Signal} {Extraction} for {Imaging} {X}-ray
  {Polarimetry}.
\newblock {\em arXiv:2107.08289 [astro-ph]}, July 2021.
\newblock arXiv: 2107.08289.

\bibitem{lakshminarayanan_simple_2017}
Lakshminarayanan, B., Pritzel, A. and Blundell, C.
\newblock Simple and scalable predictive uncertainty estimation using deep
  ensembles.
\newblock In {\em Proceedings of the 31st {International} {Conference} on
  {Neural} {Information} {Processing} {Systems}}, {NIPS}'17, pages 6405--6416,
  Long Beach, California, USA, December 2017. Curran Associates Inc.

\bibitem{fort_deep_2019}
Fort, S., Hu, H. and Lakshminarayanan, B.
\newblock Deep {Ensembles}: {A} {Loss} {Landscape} {Perspective}.
\newblock {\em arXiv:1912.02757 [cs, stat]}, December 2019.
\newblock arXiv: 1912.02757.

\bibitem{hoffmann_deep_2021}
Hoffmann, L. and Elster, C.
\newblock Deep {Ensembles} from a {Bayesian} {Perspective}.
\newblock {\em arXiv:2105.13283 [cs, stat]}, May 2021.
\newblock arXiv: 2105.13283.

\bibitem{steppa_hexagdly_2019}
Steppa, C. and Holch, T.L.
\newblock {HexagDLy} - {Processing} hexagonally sampled data with {CNNs} in
  {PyTorch}.
\newblock {\em SoftwareX}, 9:193--198, January 2019.
\newblock arXiv: 1903.01814.

\bibitem{agostinelli_geant4simulation_2003}
Agostinelli, S., Allison, J., Amako, K., Apostolakis, J., Araujo, H. et~al.
\newblock Geant4—a simulation toolkit.
\newblock {\em Nuclear Instruments and Methods in Physics Research Section A:
  Accelerators, Spectrometers, Detectors and Associated Equipment},
  506(3):250--303, July 2003.

\bibitem{kislat_analyzing_2015}
Kislat, F., Clark, B., Beilicke, M. and Krawczynski, H.
\newblock Analyzing the data from {X}-ray polarimeters with {Stokes}
  parameters.
\newblock {\em Astroparticle Physics}, 68:45--51, August 2015.

\bibitem{weisskopf_understanding_2010}
Weisskopf, M.C., Elsner, R.F. and O'Dell, S.L.
\newblock On understanding the figures of merit for detection and measurement
  of x-ray polarization.
\newblock {\em arXiv:1006.3711 [astro-ph]}, page 77320E, July 2010.
\newblock arXiv: 1006.3711.

\end{thebibliography}

\appendix

\section{Appendix}

\subsection{Epistemic error estimator}
With the epistemic uncertainties assumed to follow ${\rm VM}(0,\kappa^e_i)$; $\kappa^e_i$ can be estimated from the output of a deep ensemble with M models $\{\hat{\theta}_{ij}\}^M_{j=1}$:

\begin{equation}
    \bar{R}^2_i = \left(\frac{1}{N}\sum_{j=1}^M\cos2\hat{\theta}_{ij}\right)^2 + \left(\frac{1}{N}\sum_{j=1}^M\sin2\hat{\theta}_{ij}\right)^2 
\end{equation}
\begin{equation}
    \label{eqn:epis}
    \frac{I_1(\hat{\kappa}_i^e)}{I_0(\hat{\kappa}_i^e) } = \bar{R}_i,    
\end{equation}
with the modified Bessel functions $I_0$ and $I_1$.

\subsection{Posterior distribution on $p_0, \phi$}
\label{sec:posterior}
For a weighted polarization estimate with observations $\{\hat{\theta}\}^N_{i=1}$ and weights $\{w_i\}^N_{i=1}$ using eqs.~\ref{eqn:p}--\ref{eqn:th}, \cite{kislat_analyzing_2015} derive the joint error distribution (posterior) for the polarization fraction and EVPA estimators as
\begin{equation}
\begin{aligned}
    p(p_0,\phi| \hat{p}_0,& \hat{\phi}) ={} \frac{\sqrt{N_{\rm eff}}\hat{p}_0 \mu^2}{2\pi\sigma} \times \\
    &\exp\Bigg[-\frac{\mu^2}{4\sigma^2}\Bigg\{ \hat{p}_0^2 + p_0^2 - 2\hat{p}_0 p_0\cos(2(\hat{\phi} - \phi)) \\ & - \frac{\hat{p}_0^2p_0^2\mu^2}{2}\sin^2(2(\hat{\phi} - \phi)) \Bigg\} \Bigg],
\end{aligned}
\label{eqn:posterior}
\end{equation}
where 
\begin{equation}
    \sigma = \sqrt{\frac{1}{N_{\rm eff}} \left( 1 - \frac{p_0^2\mu^2}{2}\right)}.
    \label{eqn:sig}
\end{equation}
and 
\begin{equation}
    N_{\rm eff} = \frac{(\sum_{i=1}^N w_i)^2}{\sum_{i=1}^N w_i^2}
    \label{eqn:neff}
\end{equation}
This assumes a uniform prior over $(p_0, \phi)$. Here, $\mu$ is the empirical modulation factor for the entire set of track observations (i.e. $\mu = \hat{p}_0$ for $p_0 = 1$). $\mu$ is effectively the instrument polarization response.

Confidence intervals and the MDP$_{99}$ (\S\ref{sec:mdp99}, the 99\% upper limit for $p_0 = 0$) are derived from this posterior probability distribution.
In cases where $\mu$ and $p_0$ are not close to 0, the Gaussian approximation for the marginalized errors below is sufficient
\begin{equation}
    \sigma(p_0) \approx \sqrt{\frac{2-\hat{p}_0^2\mu^2}{(N_{\rm eff}-1)\mu^2}},
    \label{eqn:sigp}
\end{equation}
\begin{equation}
    \sigma(\phi) \approx \frac{1}{\hat{p}_0\mu\sqrt{2(N_{\rm eff}-1)}}.
    \label{eqn:sigth}
\end{equation}

High $\mu$ and high $N_{\rm eff}$ are both desirable to minimize the errors on recovered polarization parameters.  

\subsection{Minimum detectable polarization (MDP$_{99}$)}
\label{sec:mdp99}
The MDP$_{99}$ is calculated for an unpolarized $p_0 = 0$ source and is given by (using posterior eq.~\ref{eqn:posterior})
\begin{equation}
    \int_0^{\rm MDP_{99}} \int_{-\pi/2}^{\pi/2} p(p_0, \phi | \hat{p}_0, \hat{\phi})d\phi dp_0 = 0.99
\end{equation}
\begin{equation}
    {\rm MDP}_{99} = \frac{4.29}{\mu\sqrt{N_{\rm eff}}}
\end{equation}

\subsection{Maximising the SNR}
We define the signal-to-noise ratio (SNR)

\begin{equation}
    {\rm SNR} \propto 1/{\rm MDP}_{99} \propto \mu\sqrt{N_{\rm eff}}. 
\end{equation}

This is simply the inverse of the MDP$_{99}$ (without constants); an optimal weighting scheme should maximize the SNR for a fixed number of events $N$. We can expand the SNR explicitly using our weighted estimators from \textsection3,

\begin{equation}
    {\rm SNR} \propto \sqrt{\frac{\left(\sum_{i=1}^N2w_i\cos2\hat{\theta}_i\right)^2 + \left(\sum_{i=1}^N2w_i\sin2\hat{\theta}_i\right)^2}{\sum_{i=1}^Nw_i^2}}. 
\end{equation}

Expanding, squaring, and dropping constants (which do not affect maximization) we obtain
\begin{equation}
    {\rm SNR}^2 \propto \frac{\sum_{i,j,i\neq j}w_iw_j\cos2(\hat{\theta}_i - \hat{\theta_j})}{\sum_{k}w_i^2}. 
\end{equation}

The estimators $\hat{\theta}_i$ are random variables. The true values $\theta_i$, also random variables, follow the distribution eq.~\ref{eq:likelihood} (since they are perfectly known). Assuming the $\hat{\theta}_i$ are unbiased estimators of $\theta_i$ (approximately true for both moment analysis and ResNet-18s, \S\ref{sec:rec}) we can say
\begin{equation}
    \hat{\theta}_i = \theta_i + \epsilon_i
\end{equation}
where the measurement errors $\epsilon_i$ are independent random variables drawn from the same family of distributions with 
\begin{equation}
    \mathbb{E}[\epsilon_i] = 0, {\rm Var}[\epsilon_i] = \sigma^2_i.
\end{equation}

The specific distribution of the measurement errors $\epsilon_i$ will depend on the $\hat{\theta}_i$ estimation method; however since $\hat{\theta}_i$ are periodic, $\epsilon_i$ should follow a periodic distribution. For any $\epsilon_i$ distribution with the above properties, we can find the distribution for $\hat{\theta}_i$ as the convolution of the $\theta_i, \epsilon_i$ distributions
\begin{equation}
    \hat{\theta}_i \sim \frac{1}{2\pi} \big(1 + \mu_ip_0\cos[2(\hat{\theta}_i - \phi)] \big),
    \label{eqn:hat}
\end{equation}
where $0 \leq \mu_i < 1$ and $\mu_i(\sigma^2_i)$.
In other words, the distribution of estimators $\hat{\theta}_i$ are the same as the distribution of the true values $\theta_i$ but with a reduced modulation factor $\mu_i$. The measurement noise will decrease the sinusoidal modulation by a factor $\mu_i$ for the specific event $i$. 

Knowing the distributions of $\hat{\theta}_i$, eq.~\ref{eqn:hat}, we take the expectation over SNR$^2$ (dropping constant $p_0$)
\begin{equation}
    \mathbb{E}\left[{\rm SNR}^2\right] \propto \frac{\sum_{i,j,i\neq j}w_iw_j\mu_i\mu_j}{\sum_{k}w_k^2}.
\end{equation}
Finally maximizing this expression with respect to $\{w_i\}$ in the large $N$ limit, we find 
\begin{equation}
    w_i = \mu_i, 
    \label{eqn:weight}
\end{equation}
i.e., the optimal weight for an event with observation $\hat{\theta}_i$ is given by its expected $\mu$. Note for $N \sim 20$ this already holds with high accuracy; useful polarization measurements typically have $N > 1000$.

Under the assumption of von Mises distributed $\hat{\theta}$ errors
\begin{equation}
    \epsilon_i \sim {\rm VM}(0,\kappa),
\end{equation}
we have shown in \S3 that
\begin{equation}
    w_i = \mu_i = \frac{I_1(\kappa_i)}{I_0(\kappa_i)}.
\end{equation}
These weights maximize the SNR.

\begin{figure}[t!]
\centering
\includegraphics[width=0.9\textwidth]{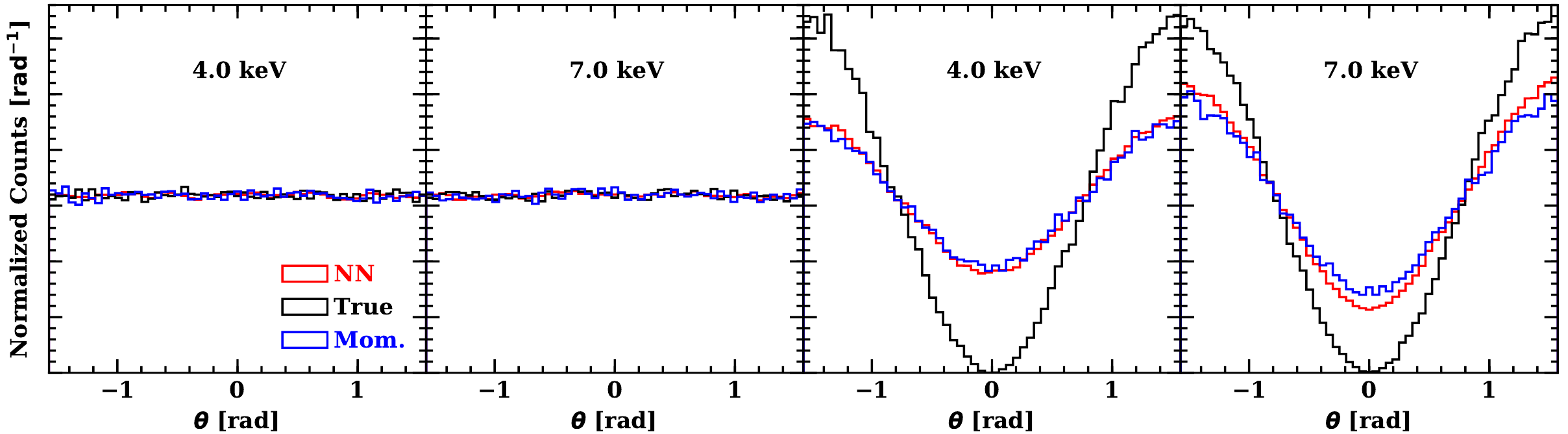}
\caption{Track angle recovery for unpolarized, $p_0 = 0$, (left two panels) and 100\% polarized, $p_0 = 1$, (right two panels) simulated data for $4.0$ and $7.0$\,keV. The true photoelectron angle distribution is shown in black; standard moment analysis reconstruction is in blue and ResNet-18 deep ensemble in red.}
\label{fig:hist}
\end{figure}
\subsection{Recovered emission angles}
\label{sec:rec}
We show the recovered $\hat{\theta}$ distribution for our deep ensemble and the moment analysis at two example energies in fig.\ref{fig:hist}. The unpolarized examples show negligible residual polarization in our method while the polarized examples suggest an increase in deep ensemble polarization signal ($\mu$), especially at higher energies. As described in the main text, the improvement in the actual polarization estimator is even greater when weighting using event uncertainties.
All plots suggest a lack of $\hat{\theta}$ prediction bias in our method.

\end{document}